Research article

# Resolution of complex fluorescence spectra of lipids and nicotinic acetylcholine receptor by multivariate analysis reveals protein-mediated effects on the receptor's immediate lipid microenvironment

Jorge J Wenz and Francisco J Barrantes*

Address: UNESCO Chair of Biophysics and Molecular Neurobiology and Instituto de Investigaciones Bioquímicas de Bahía Blanca, B8000FWB Bahía Blanca, Argentina

Email: Jorge J Wenz - jwenz@criba.edu.ar; Francisco J Barrantes* - rtfjb1@yahoo.com

* Corresponding author

## Abstract

Analysis of fluorescent spectra from complex biological systems containing various fluorescent probes with overlapping emission bands is a challenging task. Valuable information can be extracted from the full spectra, however, by using multivariate analysis (MA) of measurements at different wavelengths. We applied MA to spectral data of purified *Torpedo* nicotinic acetylcholine receptor (AChR) protein reconstituted into liposomes made up of dioleoylphosphatidic acid (DOPA) and dioleoylphosphatidylcholine (DOPC) doped with two extrinsic fluorescent probes (NBD-cholesterol/pyrene-PC). Förster resonance energy transfer (FRET) was observed between the protein and pyrene-PC and between pyrene-PC and NBD-cholesterol, leading to overlapping emission bands. Partial least squares analysis was applied to fluorescence spectra of pyrene-PC in liposomes with different DOPC/DOPA ratios, generating a model that was tested by an internal validation (leave-one-out cross-validation) and was further used to predict the apparent lipid molar ratio in AChR-containing samples. The values predicted for DOPA, the lipid with the highest $T_m$, indicate that the protein exerts a rigidifying effect on its lipid microenvironment. A similar conclusion was reached from excimer formation of pyrene-PC, a collisional-dependent phenomenon. The excimer/monomer ratio (E/M) at different DOPC/DOPA molar ratios revealed the restricted diffusion of the probe in AChR-containing samples in comparison to pure lipid samples devoid of protein. FRET from the AChR (donor) to pyrene-PC (acceptor) as a function of temperature was found to increase with increasing temperature, suggesting a shorter distance between AChR and pyrene PC. Taken together, the results obtained by MA on complex spectra indicate that the AChR rigidifies its surrounding lipid and prefers DOPA rather than DOPC in its immediate microenvironment.

**PACS Codes:** 32.50.+d, 33.50.Dq

## 1. Introduction

The nicotinic acetylcholine receptor (AChR) is an oligomeric protein deeply embedded in a lipid microenvironment that differs in physical terms from the rest of the membrane bilayer: the AChR is surrounded by boundary ("annular", "shell") lipid [1,2] that is immobilized relative to the bulk bilayer lipid in native membranes derived from *Torpedo* postsynaptic membrane. The lipid bilayer beyond the lipid annulus has also been reported to be affected, albeit to a lesser extent, by the presence of the AChR protein [reviewed in [3,4]]. The boundary lipid has a lesser degree of water penetration than the bulk lipid bilayer and has the characteristics of the liquid-order ($l_o$) lipid state [5], similar to that purported to occur in the so-called "raft" lipid domains. In reconstituted cholesterol (Chol)-containing liposomes, AChR apparently determines the lateral phase separation of certain phospholipids such as phosphatidic acid (PA), but not of others (phosphatidyl-choline, -ethanolamine or -glycerol), causing the formation of lipid domains segregated from the bulk lipid matrix [6]. Incorporation of AChR into pure phosphatidylcholine (PC) liposomes stabilizes the AChR in a nonresponsive, desensitized state, whereas reconstitution into PA- or Chol-containing liposomes favors a functional, resting state of the AChR. There is an increase in both the lateral packing density and phase transition temperature of reconstituted bilayers composed of some lipid mixtures (POPC/DOPA or POPC/POPA) but not others (POPC) [7], suggesting a differential interaction with membrane lipids.

Fluorimetric methods are extensively used to study the lipid organization in bilayers in combination with appropriate (intrinsic or extrinsic) fluorescent probes. Even after careful design and correct experimental performance, however, with suitable fluorescence spectra being recorded, the experimentalist is often frustrated by the paucity of information that can be extracted by direct examination of spectral data. Analysis of fluorescent spectra from complex biological systems containing various components and fluorescent probes with overlapping emission bands and/or interfering signals is therefore not an easy task [8]. Valuable information can be extracted from the full spectra, however, by using multivariate analysis (MA) of measurements at different wavelengths, as they allow the extraction of information that is not immediately apparent by simple visual inspection of the spectra [9]. Thanks to the availability of digitized spectroscopy data and powerful software, multivariate calibration methods are playing an increasingly important role because of their ability to exclude known or unknown interfering signals from the calculus. In complex mixtures, laborious and time- and reagent-consuming chemical separation techniques are usually needed to eliminate such interfering compounds (if at all possible). MA on the other hand can save time and effort since all that it requires is a carefully designed experiment and a reasonable amount of computing time. The analysis is carried out in several stages: a) a calibration step, in which a mathematical relationship between the spectra and the component concentration(s) is established from a set of reference samples; b) a validation step, where the performance of the regression model is tested using a set of reference samples (either the same one used to construct the model or a different one); and c) use of the validated model to determine the component concentrations in unknown samples from their whole spectra. A "weight" is assigned to each measured variable (e.g. wavelength) by attributing a coefficient in the regression model according to the variable's influence on the model and its correlation with the known dependent variable (e.g. concentration). The precision of the calculations is drastically improved by enhancing the probability of detecting experimental errors and/or anomalous samples [10-12].

In the present work we make use of Partial Least Squares (PLS-1) [13] as a multivariate tool to ascertain the influence of the AChR protein on the lipid organization and fluidity of reconstituted lipid membranes. To our knowledge, multivariate methods have so far not been employed in fluorescence studies of protein-lipid interactions in biomembranes. The initial model aimed at predicting the "apparent" molar fraction of dioleoyl-phosphatidic acid (DOPA) (and its counterpart, dioleoyl-phosphatilcholine, DOPC) in AChR-containing membranes. To this end the fluorescence emission spectra of AChR-free liposomes were initially correlated with varying molar fractions of DOPA in the mixture. The excimer to monomer ratio of pyrene-phosphatidylcholine (PyPC), which also reports on bilayer fluidity and organization [14,15], was studied next in AChR-free and receptor-containing samples as a function of DOPA molar fraction using spectral analysis of the emission bands of the two species of PyPC. Finally, the apparent Förster resonance energy transfer (FRET) from AChR to PyPC was also investigated by MA of areas in the emission spectra of donor and acceptor bands. The combined approaches yield an integrated view of the spectral properties of the complex mixtures of AChR and lipids, enabling us to extract valuable information on the interaction between the two.

## 2. Methods
### 2.1. Materials
Frozen electric organs from *Torpedo californica* were obtained from Aquatic Research Consultants (San Pedro, CA) and stored at -70°C until further use. DOPA, DOPC and Chol were obtained from Avanti Polar Lipids, Inc (Alabaster, AL). NBD-Chol and cholic acid were purchased from Sigma Chemical (St. Louis, MO). PyPC was obtained from Molecular Probes Inc. (Eugene, OR, USA). Mid-range protein molecular weight markers were purchased from Promega (Madison, USA). All other reagents were analytical quality. Lipid stock solutions were prepared in chloroform/methanol (2:1 v/v) at a concentration of 0.5 mg/ml and 0.05 mg/ml for PyPC. All solutions were stored at -20°C before use.

### 2.2. Purification of the AChR protein from AChR-enriched membranes
All steps were carried out at 4°C. AChR-enriched membranes from electric tissue of *Torpedo californica* were prepared as described previously [16]. The AChR protein was purified from these membranes by affinity chromatography using Affi-Gel 10 resin (Bio-Rad Laboratories, CA, USA) previously activated as described elsewhere [17] with slight modifications. Briefly, the affinity column was prepared by coupling cystamine to the matrix, reduction of disulphide bonds with dithiotreitol and a final activation with bromoacetylcholine bromide. Derivatization of the gel was followed qualitatively in each step by assaying for sulfhydryl groups with DTNB (5,5'-dithiobis (2-nitrobenzoic acid)). The activated gel was stored at 4°C and used within a month.

AChR-enriched membranes with a protein concentration of 6–7 mg/mL were obtained from 90 g of electric tissue and brought to a protein concentration of 2 mg/mL in MOPS buffer (10 mM MOPS, 0.1 mM EDTA, 100 mM NaCl, 0.02% sodium azide (w/v), pH 7.4). Membranes were solubilized by adding cholic acid to give a final concentration of 1% (w/v) with gentle stirring for 20 min. The detergent extract was centrifuged in a Ti 70 rotor at 34,000 rpm for 45 min. Ten ml of the activated affinity column were washed with 4–5 volumes of water followed by 2 volumes of MOPS buffer containing 1% cholate, and the supernatant was applied to the column. To exchange native lipids for the desired exogenous lipid (reconstitution), the column was pre-equilibrated with the following sequence of DOPC solutions in MOPS buffer-1% cholate: 4 vol of 0.1 mg/mL, 2.5 vol of 2 mg/mL and 3 vol of 0.1 mg/mL. The AChR was eluted by applying 3 vol of MOPS buffer-1% cholate, 10 mM carbamoylcholine chloride and 0.1 mg/mL DOPA or DOPC. Protein concentration was monitored by measuring $A_{280}$ and fractions with values greater than 0.1 were pooled. To eliminate cholate and carbamoylcholine, and in order to form vesicles, the mixture was dialyzed using 12,000 MW cut-off cellulose membranes at 4°C against 200 volumes of MOPS buffer, with gentle stirring for 40 h and buffer replacement every 8 h.

### 2.3. Characterization of purified AChR

To verify the purity of the AChR protein, SDS-gel electrophoresis was performed under denaturing conditions [18] using 0.75 mm thick gels and 10% total acrylamide concentration. Electrophoresis was carried out in a Mini Protean II Cell (BioRad laboratories, Inc., CA, USA) for 3 h at 15 mA, supplied by a Power Pack 3000 (BioRad). AChR subunits were visualized using a silver staining kit (Amershan, Uppsala, Sweden) and identified by comparison with mid-range protein molecular weight markers.

Protein concentration was assayed according to [19]. Lipids were extracted as described in [20] and assayed for inorganic phosphorus content according to [21]. Total lipid concentration was 0.12 mg/mL (0.163 mM).

### 2.4. Lipid suspension and re-reconstitution of purified AChR

Lipid suspensions were prepared by mixing the appropriate amount of stock solution of each component to obtain mixtures of DOPA/DOPC with different molar ratios, ranging from pure DOPA to pure DOPC but always totaling 60 mol%. NBD-Chol and PyPC concentrations were maintained at 40% and 5%, respectively; the latter was computed over the final lipid concentration of each sample: 40 μM (see Table 1). Taking into account the lipids contained in the purified AChR preparation, the lipid to protein ratio was calculated and adjusted in order to achieve the desired lipid and protein concentrations upon subsequent incorporation of purified AChR into the samples.

The lipid mixtures were dried under nitrogen at room temperature while rotating the round-bottomed tubes to form the lipid film. NBD-Chol was highly adherent to glass tubes and almost insoluble in MOPS buffer containing 1% cholate. Heating, sonication and/or vortexing for more

**Table 1: Composition of liposomes used in fluorescence experiments.**

| | Mol % | | | | | | | | | |
|---|---|---|---|---|---|---|---|---|---|---|
| Sample No.: | 1 | 2 | 3 | 4 | 5 | 6 | 7 | 8 | 9 | 10 |
| PyPC | 0 | 0 | 5 | | | | | | | |
| DOPA | 50 | 20 | 20 | 60 | 50 | 40 | 30 | 20 | 10 | 0 |
| DOPC | 50 | 40 | 40 | 0 | 10 | 20 | 30 | 40 | 50 | 60 |
| NBD-Chol | 0 | | 40 | | | | | | | |

Samples 1 and 2 are control liposomes; sample 3 corresponds to the AChR-containing liposomes, with a lipid to protein molar ratio of 500:1. Excluding PyPC (5 mol% extra), final lipid concentration was 40 μM for all samples, amounting to 100%. Except for sample 2, all other samples were assayed in duplicate.

than half an hour did not improve the solubilization of the lipids, nor did increasing the cholic acid content up to 25% (w/v). To circumvent this problem, the dried film was resuspended in 7.5 μL methanol with vortexing for 2 min, followed by addition of MOPS buffer containing 1% cholate to reach a final volume of 1.5 mL; the final methanol concentration was thus 0.5% (v/v). Since DOPA and DOPC have low temperatures (-8 and -20°C, respectively) this step could be accomplished at room temperature (i.e. a temperature higher than that of the highest $T_m$ of the constitutive lipids). At this stage, the total lipid concentration of the suspension was 0.5 mg/mL. Samples were then vigorously vortexed for 2 min.

The purification procedure yielded membranes with a lipid to protein ratio of 45:1 (mol:mol). Purified AChR protein (0.2 mg/mL) was added to aliquots of the suspension, yielding re-reconstituted membranes with a lipid to protein ratio of 500:1 (mol:mol) and a DOPA/DOPC/NBD-Chol molar ratio of 20/40/40, keeping the total lipid concentration at 40 μM (Table 1). In order to eliminate cholate and form vesicles, the mixtures were dialyzed using 12,000 MW cut-off cellulose membranes at 4°C against 200 volumes of MOPS buffer, with gentle stirring for 40 h and buffer replacement every 12 h.

### *2.5. Fluorescence measurements*

Fluorescence emission spectra were recorded with an SLM 4800 ISC spectrofluorimeter (SLM Instruments, Urbana, IL) with a light beam from a Hannovia 200 W mercury/xenon arc. Measurements were carried out in 5 × 5 mm quartz cuvettes in the temperature range of 25–55°C. Optimal excitation wavelengths for AChR, PyPC and NBD-Chol were 290, 340 and 478 nm, respectively. Fluorescence emission spectra were obtained 15 min after reaching the desired working temperature, which was regulated by a thermostated circulating water bath (Haake, Darmstadt, Germany) with a heating rate of 0.5°/min monitored inside the cuvette with a thermocouple. Background fluorescence was measured in samples made up of pure DOPA/DOPC (50:50 mol/mol) mixtures, and subtracted from probe-containing samples. The fluorescence intensity of these background samples amounted to less than 5–10% of that of the probe-containing counterpart. In FRET experiments samples were excited at 290 nm and fluorescence emission spectra of AChR ($F_{330nm}$), PyPC ($F_{374nm}$ and $F_{474nm}$) and NBD ($F_{534nm}$) were recorded.

### 2.6. Multivariate analysis of fluorescence spectra

Univariate calibration is when one response of e.g. an instrument (absorbance, fluorescence, chromatographic peak, etc.) is related to the concentration of a single component. Multivariate calibration, on the other hand, relates multiple measurements and/or responses of an instrument (i.e. fluorescence intensity at several wavelengths, spectrum) to a property or properties of a sample. This offers several advantages over univariate approaches: a) information from multiple measured variables can be extracted and exploited; b) it is possible to analyze multiple components simultaneously, which is difficult with univariate analysis; c) significant variables can be selected by weighing their importance, discarding those that are uncorrelated with a given dependent variable; d) the use of multiple variables can also improve precision in prediction, since statistics demonstrate that repeating a measurement $n$ times and calculating a mean value will result in a factor of $\sqrt{n}$ reduction in the standard deviation of the mean; and e) it enhances the detection of errors, interferences and/or anomalous samples, given that the likelihood that these failures affect the response in all the variables decreases as the number of variables increases [10,11]. One of the advantages of "inverse" methods of multivariate calibration and prediction (MLR, Multiple Linear Regression; PCR, Principal Component Regression and PLS-1, Partial Least Squares) over "classical" methods (DLCS, Direct Classical Least Squares and ICLS, Indirect Classical Least Squares) is their ability to generate good predictive models in complex systems with known and/or unknown components [10]. Their ability to implicitly model all sources of variation means that they can perform appropriately even if additional, chemical or physical sources of variation such as interferences are present in the samples. The proper functioning of such models requires that the measurements adequately differentiate the component of interest from other sources of variance and that the measurements and/or instrument responses vary in a sufficiently linear manner with the concentration. In practice, this means that much less explicit knowledge of the system and potential interferences is required, but that the calibration step may involve more work than when using "classical" methods.

Among the multivariate approaches, PLS-1 is undoubtedly the most widely used method for constructing mathematical models able to predict properties (e.g. concentration of a compound) of unknown samples given an instrument output (e.g. spectrum). The a priori calibration step relating known concentrations of a compound in reference samples with their full spectra permits the building of a mathematical model relating the spectral data to the concentration of the compound. This model can then be used to predict the concentration of the compound in unknown samples on the basis of their spectra.

### 3. Results
#### 3.1. Pre-processing of the spectra and selection of appropriate wavelengths

"Pre-processing" commonly refers to any mathematical manipulation of data prior to primary analysis (e.g. PLS-1 in the present study) and is used to remove or reduce irrelevant sources of variation, random or systematic. It can operate on one variable at a time over all samples, or on one sample at a time over all variables (the present case). Here the operational definition of "pre-processing" refers to the smoothing and normalization of all samples at all wavelengths, followed by a selection of the appropriate interval. The pre-processed data were subsequently used in the "processing" step proper, i.e. PLS-1 analysis. Fig. 1A shows the raw fluorescence emission spectra of AChR-free liposomes with varying DOPA and DOPC molar fractions (0–60%) and AChR-containing liposomes with a DOPA and DOPC molar fraction of 20 and 40%, respectively, at the excitation wavelength of PyPC (340 nm).

Even though the fluorescent probe concentration was kept constant (see Table 1), noticeable differences in the fluorescence intensity were apparent at all wavelengths in raw spectra (Fig. 1A). Direct examination of the spectral data did not provide conclusive information, which is not surprising in view of the complex nature of the biological samples.

In order to eliminate experimental and instrumental noise of the recorded spectra, the data were initially smoothed using the Savitsky-Golay procedure, with a window width of 8 points (i.e. wavelengths). The middle point of the interval was then estimated using a second-order polynomial fit, whereby four points at the beginning and the end of each spectrum were left out by the averaging procedure. The spectra were subsequently normalized to their highest intensity value, i.e. to the first peak (374 nm) of the PyPC monomer band, which spans an area between 358 and 434 nm. The resulting data (Fig. 1B) showed more clearly the section of the spectrum where the most noticeable changes in fluorescence intensity occurred: a restricted region (434–590 nm, rectangle in Fig. 1B) corresponding to the PyPC excimer (first band, ~474 nm) and the fluorescent Chol analogue NBD-Chol (second band, ~534 nm) emission bands. The spectral window chosen reports on the changing DOPA molar fraction (and hence on the DOPC molar fraction). NBD-Chol acts as an energy transfer acceptor of the PyPC donor (NBD-Chol excitation wavelength, 460 nm). The reference spectra of the individual probes are shown in Additional file 1.

### 3.2. Constructing the model

After selecting the useful spectral region, PLS-1 was used to construct a model relating the DOPA molar fraction of 14 AChR-free samples to their spectra [22]. Though in this instance the DOPA molar fraction was chosen as a variable, the DOPC molar fraction should in fact yield similar results since the two are inversely correlated and the sum of their molar fractions amounts to 60% in all samples. The pre-selected 79 x-variables were thus the fluorescence intensity at each wavelength, collected every 2 nm in the chosen interval (434–590 nm). The model was constructed by using 1 to 6 factors (also called principal components, "PC"), computed in a data matrix of 14 samples × 79 variables. Since a requisite for inverse methods is that the number of factors calculated must be less than or equal to the number of calibration samples or the number of variables, an internal validation using a "leave-one-out cross-validation" procedure was performed in order to test the model. Here, one sample was left out, a model with a given number of factors was calculated using the remaining samples, and the residual of the left-out sample was calculated after predicting its value with the model. This was repeated for each sample and for models with 1 to 6 factors. Because of its high residual (i.e. "outlier" value), one sample was removed and 13 samples were then used for model building. The optimal number of factors was found to be 4 on the basis of the lowest RMSEP and highest explained y-variance values (see Additional file 2).

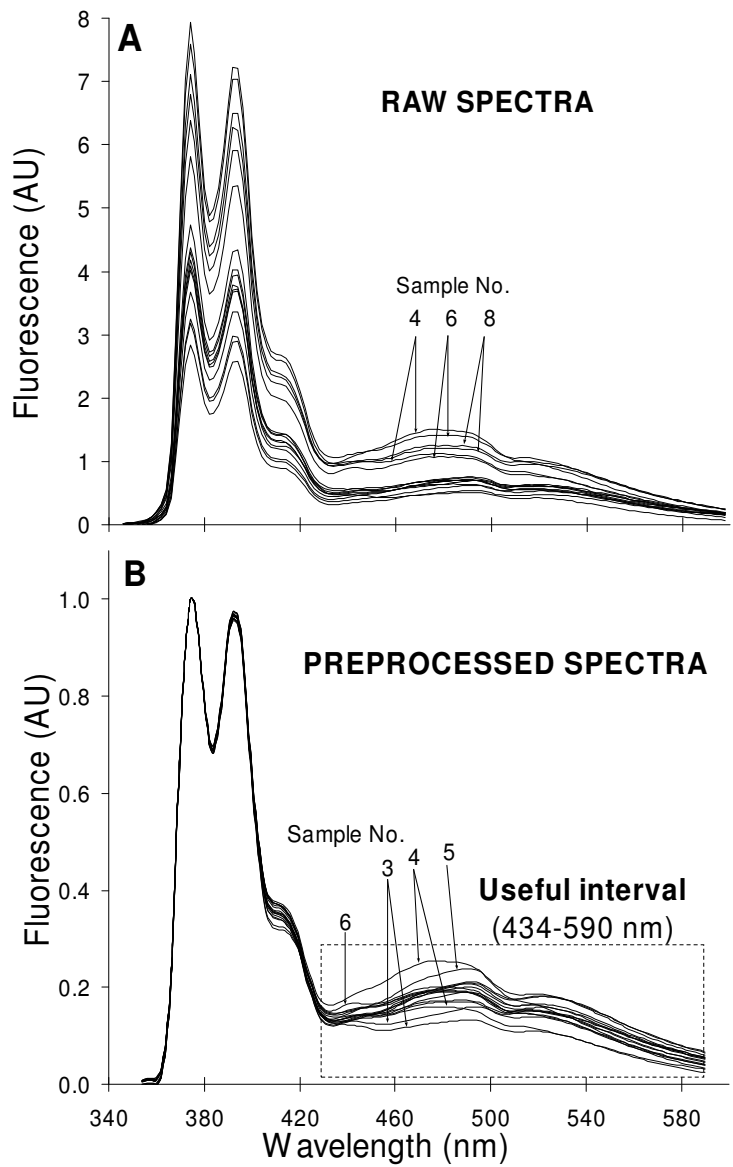

**Figure 1**
**Pre-proccesing of spectra and selection of appropriate wavelengths. A)** Raw fluorescence emission spectra of various AChR-free liposomes with varying molar ratios of DOPA/DOPC (ranging from 0 to 60%) and two AChR-containing liposomes with DOPA/DOPC (20:40). Two replicas from independent experiments are shown. In all samples, NBD-Chol and PyPC concentrations were kept at 40 and 5%, respectively. The final lipid concentration and the lipid to protein molar ratio in the AChR-containing liposomes were 40 μM and 500:1, respectively. Excitation wavelength = 340 nm. Temperature = 25°C. **B)** Pre-processed spectra obtained by smoothing throughout the Savitsky-Golay procedure (with a window width of 8 points and a second order polynomial fit) and normalizing to the highest value of each spectrum (corresponding to the first PyPC emission band at 374 nm). The wavelengths further employed for the multivariate analysis are outlined in the dotted rectangle. It is clear that the sixteen spectra in each panel are difficult to identify due to their high degree of overlap; this is precisely what multivariate analysis dissects so effectively. Where possible, spectra have been identified with the sample numbers appearing in Table 1.

### 3.3. "Apparent" DOPA and DOPC molar fraction in AChR-containing samples

The model constructed with 4 factors was used next to evaluate the apparent DOPA molar fraction in AChR-containing liposomes with a known DOPA molar fraction of 20%. The predicted DOPA molar fraction for all samples at 25°C versus the corresponding known values is shown in Fig. 2A together with their corresponding deviation, which reflects the degree of similarity between the predicted values and the calibration samples used for building the model. For each sample the deviation (95% confidence interval) is computed as a function of the sample's leverage and its x-residual variance. The higher the similarity, the smaller the deviation. A high correlation was found between known and predicted DOPA molar fractions in the AChR-free samples ($r^2$ = 0.991), reinforcing the validity of the model.

For the two AChR-containing samples assayed, the predicted DOPA fraction was considerably higher (64 and 82%) than its actual value (20%). DOPA has a higher transition temperature (-8°C) than DOPC (-20°C). This implies that the AChR rigidifies its lipid microenvironment: the observed bilayer packing is as high as that corresponding to the predicted DOPA molar fraction. The portion of the spectra employed to construct the model (434–590 nm) matches the emission band of the PyPC excimer (see Fig. 3 below), which informs on membrane fluidity. The variability found in this portion of the spectra after normalization to the highest value (first peak of the PyPC monomer) reports on excimer formation, related to membrane fluidity.

PLS-1 was applied next to the raw (Fig. 1A) and smoothed, normalized spectra (Fig. 1B) of AChR-free samples in the *entire* wavelength interval (354–590 nm). The model obtained from the raw data performed poorly (Fig. 2B) compared to that corresponding to the selected region of high variability as revealed by smoothing and normalization of the spectra. In statistical terms, this poorer performance was reflected in: a) a higher number of factors (5); b) a higher prediction error (RMSEP = 0.123); c) a lower percentage of the y-variance explained (67.6%) and d) a worse correlation between the predicted and known DOPA molar fraction, as deduced from both the lower $R^2$ (0.920) and a slope far from unity (0.919). Upon performing the first two pre-processing steps (i.e. smoothing followed by normalization), PLS-1 was applied again to the entire spectra (354–590 nm). The model obtained at this stage exhibited improved performance (Fig. 2C), as can be inferred from the statistical parameters (number of factors = 4; RMSEP, 0.060;% y-variance explained, 92.9; $R^2$ = 0.974; slope, 0.975). Finally, by using only a restricted portion of the smoothed and normalized spectra (434–590 nm) where the variability between samples is at its highest, a model with higher prediction capability could be constructed (Fig. 2A). Though a similar pre-processing and statistical treatment was also applied to the emission spectra (484–590

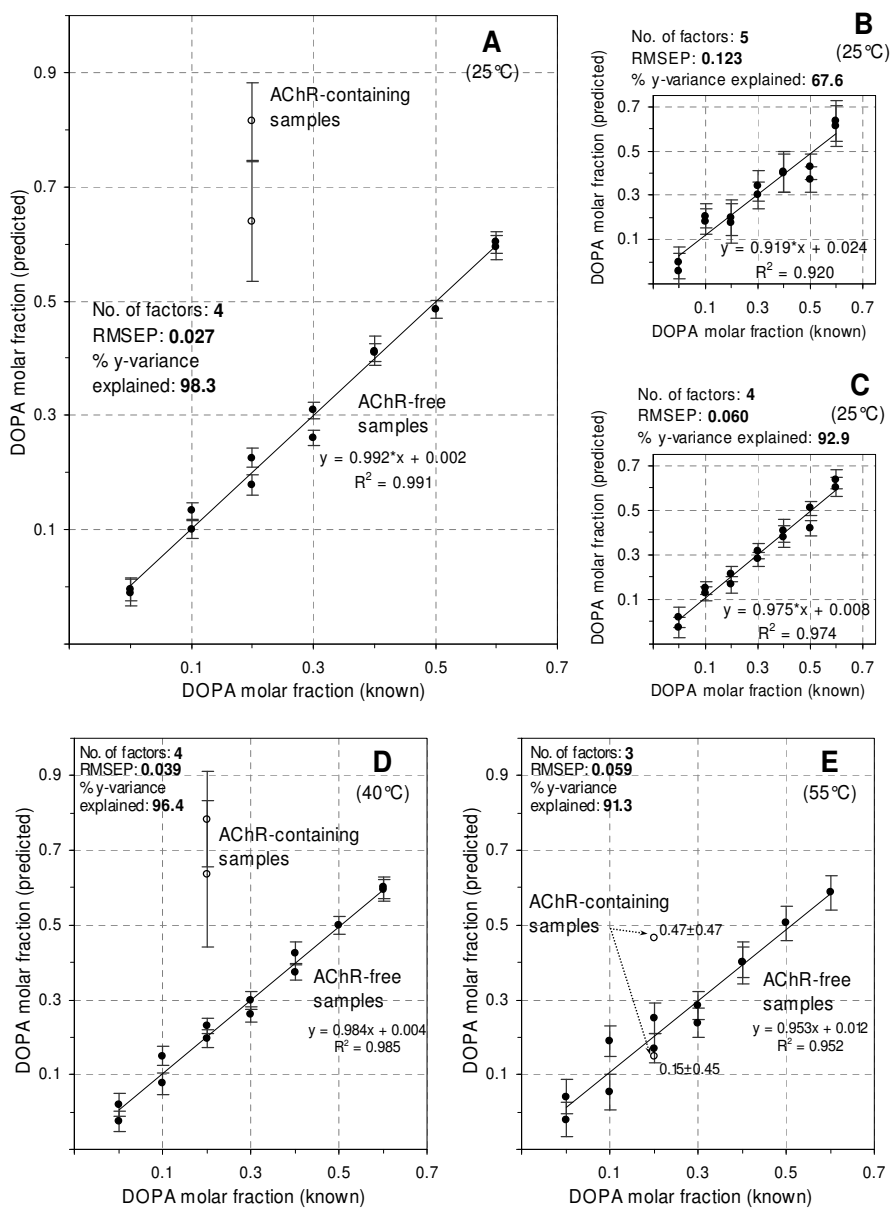

**Figure 2**
**Predicted versus known DOPA molar fraction. A-C)** Comparison of the predictive performance of PLS-1 models constructed from either pre-processed or unprocessed spectra from samples at 25°C. **A)** Model obtained after smoothing and normalizing the full spectra, followed by a selection of a restricted portion of wavelengths (434–590 nm, Fig. 1B). **B)** Model constructed with the raw and full spectra (354–590 nm, Fig. 1A). **C)** Model constructed with the smoothed, normalized and full spectra (354–590 nm). Statistical parameters, reflecting the model's performance, are shown in each panel (see text for details). Each model was tested by leave-one-out cross-validation. Since the model shown in (A) displayed the highest predictive performance, it was used to estimate the DOPA molar fraction in AChR-containing samples at 25°C. Models obtained at 40°C (panel D) and 55°C (panel E) from pre-processed spectra. Bars represent deviation, expressing the degree of similarity between the predicted samples and the calibration samples used to build each model. Deviation (95% confidence interval around the predicted value) is computed as a function of the sample's leverage and its x-residual variance. After removing outliers, linear equations in each plot were computed using the remaining numbers of AChR-free samples: 13 in (A), 14 in (B) and (C), 13 in (D) and 12 in (E).

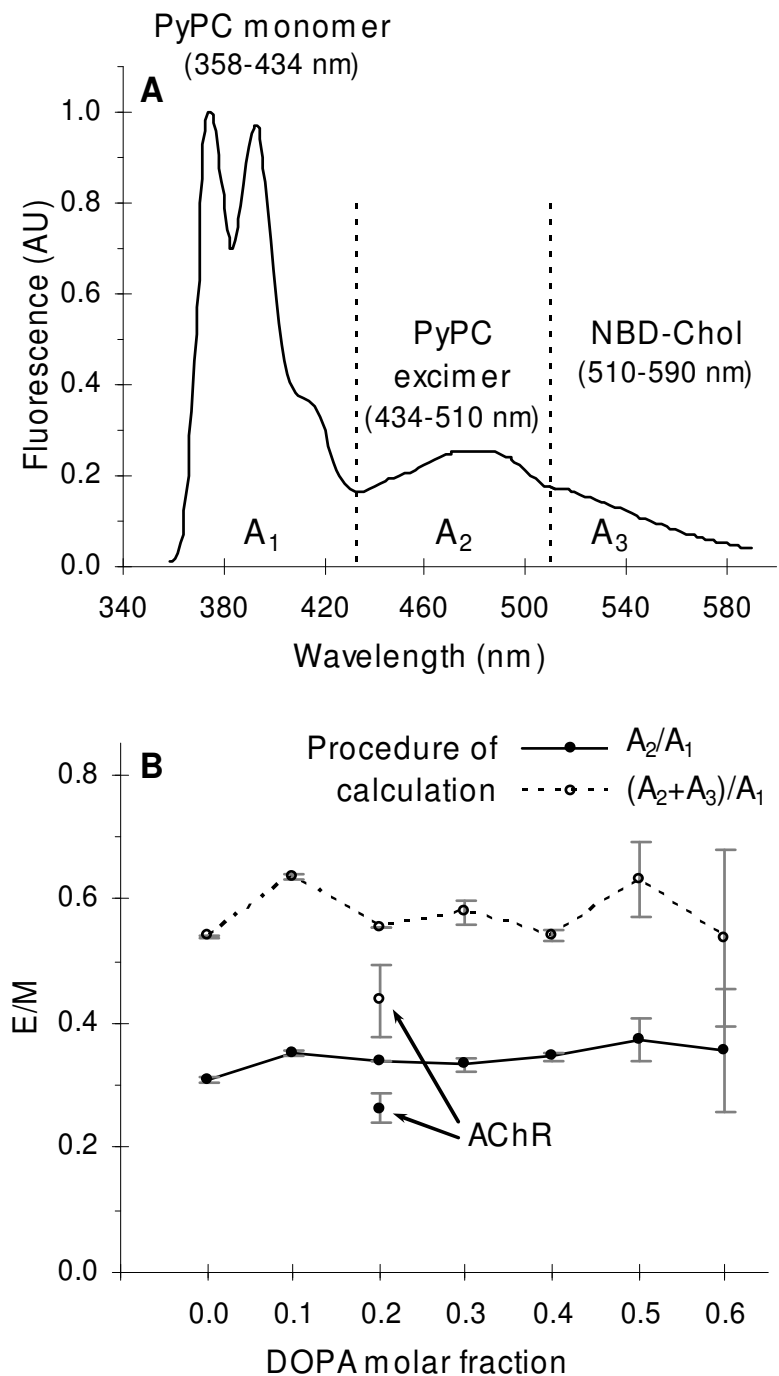

#### Figure 3
**Excimer to monomer ratio of PyPC**. A) Example of a smoothed and normalized fluorescence emission spectrum showing the areas ($A_1$, $A_2$ and $A_3$) employed to evaluate the ratio between the excimer and monomer of PyPC (E/M). Dotted lines indicate the average boundary wavelengths at which the absolute value of the first derivative of the spectrum reached a minimum. Temperature and excitation wavelength were 25°C and 340 nm, respectively. B) Estimated E/M ratios by relating the areas of the spectra using one of the two procedures shown in the legend. Error bars represent the S.D. of two independent samples. Arrows indicate the E/M values for the AChR-containing sample according to each procedure of calculation.

nm) of NBD-Chol samples excited at 478 nm, no conclusive information could be obtained by MA.

The apparent DOPA molar fraction in AChR-containing samples was investigated at higher temperatures by applying PLS-1 to spectra at 40 and 55°C (Fig. 2D and 2E). The predicted DOPA fraction in samples at 40°C was higher (64 and 78%) than its actual value (20%), indicating that AChR-phospholipid interactions, and hence the rigidifying effect of the protein on the lipid matrix, remains even at this temperature (Fig. 2D). Upon increasing the temperature to 55°C the predicted DOPA molar fraction (Fig. 2E) showed minor differences in comparison to known values (15 and 47% versus 20%). Although the model's performance (RMSEP, 0.059;% y-variance explained, 91.3; $R^2$ = 0.952; slope, 0.953) was not as good as with the lower temperature samples and higher deviations were apparent in the predicted values in AChR-containing samples, this finding suggests that interactions between the protein and its surrounding lipid matrix decrease at such high temperature.

### *3.4. PyPC excimer to monomer fluorescence ratio (E/M)*

We also investigated lipid organization and fluidity in reconstituted membranes by using PyPC, a lipid probe having the fluorophore attached at the end of the 10-C acyl chain. This fluorescent lipid partitions favorably into lipid-fluid phases [14,15] and tends to form probe-enriched domains ("patches") laterally segregated from lipids in the gel state [23]. PyPC displays a phase behavior similar to that of an unsaturated lipid with a low solubility in gel-state bilayers, which is usually attributed to poor fitting of the pyrene moiety into tightly-packed lipids in the gel state [14]. A very useful trait of pyrene probes, PyPC included, is that they form excimers when a molecule in the excited state collides with a molecule in the ground state. This phenomenon depends on both the concentration and lateral diffusion of the probe in the lipid matrix [24,14,25,15]. Experimental measurement of E/M can therefore be effectively used to sense changes in the physical state and fluidity of the bilayer. E/M may increase by a local (or apparent) augmented concentration of the probe owing to lateral segregation from the gel state into probe-enriched domains (clusters), and/or by an enhanced diffusion rate (fluidity).

E/M is usually calculated from the ratio between the fluorescence intensity at the maximum emission band of the excimer (~424 nm) and that of the monomer (~374 nm). Monomer and excimer emission bands are dissected in Additional file 1, in which reference spectra of the probe are shown. We employ the full spectral information to estimate E/M, instead of two wavelengths only, by calculating the ratio between the areas under the excimer and the monomer emission bands ($A_2/A_1$ in Fig. 3A) in the smoothed and normalized spectra. The two vertical dotted lines in Fig. 3A denote boundary points between regions, ascertained by averaging wavelengths of spectra where the first derivative was closest to or equal to zero. We also estimated E/M from the ratio between areas below the PyPC excimer + NBD-Chol and monomer emission bands $(A_2+A_3)/A_1$ to account for possible resonance energy transfer to the latter probe (Fig. 3A).

No significant variation in tendency was observed in E/M in AChR-free bilayers with changing DOPA molar fraction using either calculation procedure (Fig. 3B), suggesting that no appreciable lipid phase segregation and/or changes in fluidity occurred for any DOPA/DOPC molar ratio. This is not surprising since both DOPA and DOPC are in the fluid state at 25°C ($T_m$ of DOPA and DOPC are -8 and -20°C, respectively). The other factor possibly affecting E/M, i.e. fluidity, will have remained constant, since temperature was kept constant for all DOPA/DOPC combinations. Furthermore, PyPC did not partition or segregate into clusters and was most likely homogeneously distributed, i.e. no coexistence of solid and liquid phases took place in DOPA/DOPC bilayers at the working temperature.

When the AChR was incorporated into DOPA/DOPC (20/40 molar ratio) bilayers, a lower value of E/M was found as compared to AChR-free samples with any of the lipid combinations tested and using any of the two calibration procedures (Fig. 3B). The drop in E/M can be explained in terms of a lower collisional rate of PyPC because of the reduced diffusion rate/fluidity. Again, in view of the low $T_m$ values of the constitutive lipids it is unlikely that liquid and solid lipid phases can form and coexist at 25°C owing to the presence of the protein. The restricted diffusion of PyPC is probably due to a higher lateral packing density and a lower fluidity of the lipid matrix caused by the presence of the AChR, in agreement with the results obtained for the apparent DOPA molar fraction in AChR-containing samples.

### *3.5. Förster energy transfer (FRET) from AChR to PyPC*

The good spectral overlap between the fluorescence emission of Trp residues in the AChR protein ($F_{330}$) [26,27] and PyPC absorption bands ($F_{340}$) enabled us to employ FRET conditions by exciting samples at the protein main absorption band (290 nm). Individual spectra of AChR and PyPC are provided in Additional file 1. To investigate the effect of temperature on lipid organization within Förster's distances from the protein (reported to be ~28 Å, ref. [28]), we estimated the apparent FRET between the protein and the probe at 25, 40 and 55°C in AChR-containing samples with a DOPA/DOPC/NBD-Chol 20:40:40 molar fraction. Instead of estimating FRET by measuring the fluorescence decrease of the donor at one wavelength only, we once again exploited the multivariate spectral information by estimating FRET = $1-(A_1/A_2)$, where $A_1$ and $A_2$ are the areas below the donor (AChR) and acceptor (PyPC) emission bands as shown in Fig. 4. Since a ratio ($A_1/A_2$) is used, smoothed spectra yield the same results as those from smoothed + normalized spectra. $A_1$ and $A_2$ were computed from 304 to 360 nm and from 360 to 544 nm, respectively. The latter interval was extended up to 544 nm to account for the overlapping emission bands of fluorophores, changes in PyPC excimer:monomer ratio, and FRET occurring between PyPC excimer and NBD-Chol. By computing this extended wavelength interval for $A_2$, and not that of the monomer band only (360–434 nm), all possible fluorescence signals beyond the PyPC monomer are accounted for.

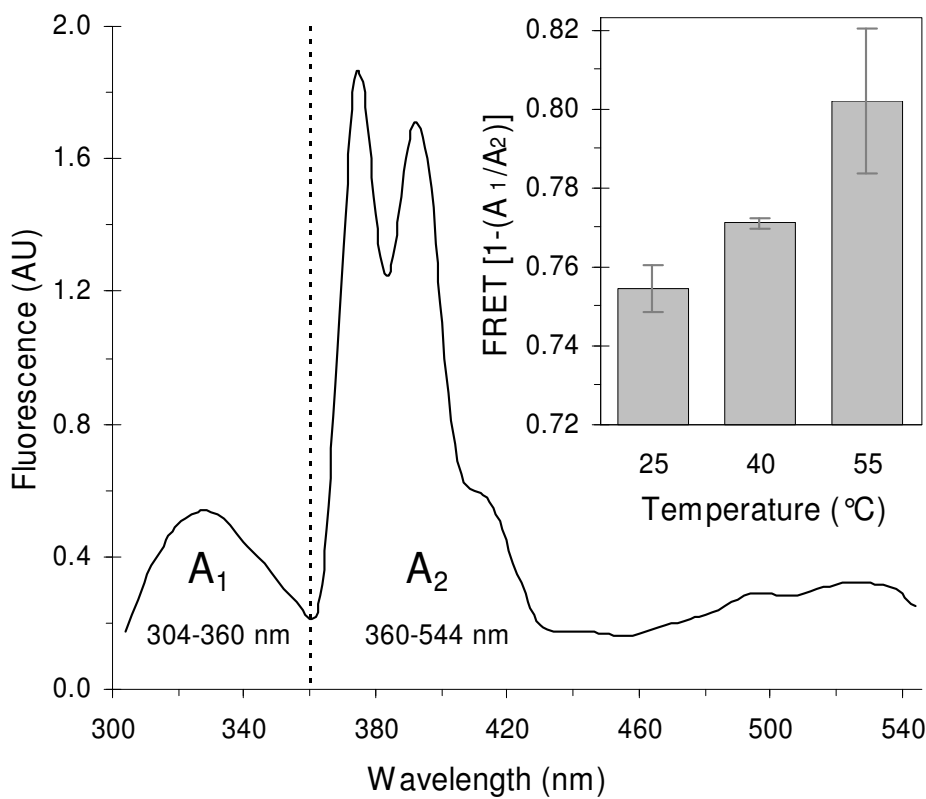

**Figure 4**
**Förster resonance energy transfer (FRET) from the AChR**. Smoothened fluorescence emission spectrum of an AChR PyPC-containing sample at 25°C showing the areas (A1 and A2) used to evaluate FRET. The dotted line indicates the average boundary wavelength at which the absolute value of the first derivative of the spectrum passed through a minimum. Excitation wavelength = 290 nm. Inset: FRET from AChR to PyPC as a function of temperature. FRET was calculated according to [$1-(A_1/A_2)$]. Liposomes contained DOPA/DOPC/NBD-Chol at a molar fraction 20/40/40 and a lipid to protein molar ratio of 500:1. PyPC concentration was 5 mol%. Error bars represent the S.D. of two independent samples.

FRET from the AChR to PyPC increased slightly upon increasing the temperature from 25 to 55°C (inset in Fig. 4), suggesting a shorter distance between the FRET partners, i.e. a closer AChR-PyPC contact.

In order to check that the increase in FRET = $1-(A_1/A_2)$ is due to an energy transfer process and not to some other phenomenon, spectra of AChR-only and PyPC-only samples were measured at the same temperatures (25, 40 and 55°C). Both $A_1$ and $A_2$ were found to diminish with temperature but not at the same rate, yielding increasing values of $A_1/A_2$ (see Additional file 3) that translate into lower [$1-(A_1/A_2)$] values in the control samples as temperature increases. These findings validate the procedure for calculating FRET and confirm that energy transfer increases with increasing temperature.

## 4. Discussion

The crosstalk between the AChR and its lipid microenvironment was investigated by applying chemometric tools to fluorescence spectra, an approach to our knowledge not previously used to study lipid-AChR interactions. This powerful statistical approach (PLS-1) enabled us to extract otherwise masked spectral fluorescence data on protein-lipid interactions under conditions known to affect the functionality of the former [reviewed in [4]].

The PLS-1 model predicted higher DOPA molar fractions (here termed "apparent") than their actual values in lipid bilayers containing the AChR. These predictions emerge from the MA applied over a selected portion of the analyzed spectra corresponding principally to the PyPC excimer band, which reports on membrane fluidity. The obtained information proved extremely difficult to extract from the raw overlapping spectra (Figs. 1A–B) by utilizing the maximum fluorescence intensity at one wavelength only. MA enabled us to detect and correlate variations across the entire spectra, establishing relationships between those wavelengths that displayed a linear-like correlation with the DOPA molar fraction. The model predicted a more rigid lipid matrix in the presence of the AChR, on the basis of the higher apparent molar fraction of the phospholipid with the highest $T_m$ (DOPA).

The series of experiments using E/M in bilayers lacking the AChR revealed no appreciable changes in the diffusion rate/fluidity with changing DOPA/DOPC molar ratios (Fig. 3B). This is not unexpected, since at 25°C both lipids are in the fluid state. Since the experiment was performed above 30°C (i.e above the highest $T_m$ of each lipid; -8° for DOPA) it is also unlikely that lateral phase segregation occurred at any of the assayed lipid molar ratios. Conversely, the drop in the E/M upon incorporation of the AChR indicates that the protein decreases the diffusion rate/fluidity of its surrounding lipid matrix, in agreement with the results of the apparent DOPA molar fraction experiments. E/M was also found to decrease in studies of AChR reconstituted in a single-phase DPPC bilayer and in coexisting liquid-solid bilayers composed of POPA/POPC[28].

AChR-PyPC FRET experiments in DOPA/DOPC bilayers revealed the preference of the AChR protein for DOPA over DOPC, in agreement with previous results showing favored interactions between PA and the AChR relative to PC, PG and PS [6,7,30,29]. The AChR is included in PA-enriched regions, separated from the bulk lipids [6,7]. We hypothesized that with increasing temperature DOPA molecules separate from the protein, whereas the opposite occurs for DOPC molecules, which move closer to the AChR. Owing to the augmented number of DOPC molecules near the protein, FRET increases with increasing temperature because a higher number of PyPC molecules (with higher affinity for DOPC) come closer to the protein. In the case of lipid mixtures exhibiting liquid-solid phase coexistence, i.e. having lipids with $T_m$'s above and below the working temperature, this enhanced affinity could lead to segregation of PA-enriched domains from the bulk lipid. In a liquid system such as the one used in this work, where no liquid-solid phase segregation is likely to occur, the minor increase in FRET with increasing temperature can most economically be interpreted in terms of a migration and re-distribution of the probe. The mixing of lipids upon heating, and the resulting diffusion of DOPC to regions near the AChR were mirrored by the diffusion of PyPC, which partitions favorably in DOPC-enriched domains. Thus, increasing temperature causes a dispersion of DOPA and a concurrent enrichment of DOPC and PyPC molecules in the immediate vicinity of the AChR. AChR incorporation into PA-containing membranes leads to an increase in both the lateral packing densities and the $T_m$'s of POPC/DOPA or POPC/POPA mixtures, an effect which is absent in pure POPC bilayers [7,30]. These changes are accompanied by an increase in the percentage of non-hydrogen-bonded lipid ester carbonyls, reinforcing the proposal that that the AChR reduces water penetration at the polar head group-acyl chain interface [5]. In the absence of the AChR protein, ideal mixing of membrane components takes place, whereas the presence of the protein leads to lateral phase segregation of certain lipids (DMPA) but not others (DMPC, DMPG, DMPE) [7,31]. The promotion of lateral segregation of POPA from a liquid phase enriched in lipids having identical acyl chains, such as POPC, leads to a tighter lipid packing and to an increase in the $T_m$'s of the lipid system. PA is required for proper receptor function [32,33], and its $T_m$ is higher than that of non-functional bilayers composed of PC only [7].

The ability of the AChR to selectively modulate the physical properties of the bilayer could stem essentially from a hydrophobic mismatch between the AChR and the fatty acyl chains and/or polar head groups. Such mismatch could also play an important role in the modulation of AChR function. The present results, obtained by application of MA to complex spectra obtained under various fluorescence spectroscopy modalities (steady-state spectra, FRET, excimer/monomer measurements) and exploiting the intrinsic fluorescence of the AChR protein or extrinsic fluorescent derivatives of phosphatidylcholine or cholesterol, support the more general notion that the AChR organizes its immediate environment with increased lateral packing density and rigidity. The functional counterpart of this finding is that ordered lipid membranes stabilize the receptor in a resting, activatable state, whereas fluid or disordered membranes favor a desensitized conformation of the receptor protein.

**Abbreviations**
MA: multivariate analysis; AChR: acetylcholine receptor; Chol: cholesterol; DOPA: dioleoylphosphatidic acid; DOPC: dioleoylphosphatidylcholine; DPPC: dipalmitoylphosphatidylcholine; DTNB: 5,5'-dithiobis (2-nitrobenzoic acid); FRET: Förster resonance energy transfer; NBD-Chol: (22-(N-(7-nitrobenz-2-oxa-1,3-diazol-4-yl)amino)-23,24-bisnor-5-cholen-3-ol; PA: phosphatidic acid; POPC: 1-palmitoyl-2-oleoylphosphatidylcholine; PC: phosphatidylcholine; POPA: 1-palmitoyl-2-oleoylphosphatidic acid; PyPC: 1-hexadecanoyl-2-(1-pyrenedecanoyl)-sn-glycero-3-phosphocholine; RMSEP: root mean square error of prediction; $T_m$: transition temperature.

## Additional material

### Additional File 1
*Reference spectra of AChR, PyPC and NBD-Chol.* Fig. 1 and experimental details.
Click here for file
[http://www.biomedcentral.com/content/supplementary/1757-5036-1-6-S1.pdf]

### Additional File 2
*REMSEP and explained y-variance.* Fig. 2 and extra information about these parameters.
Click here for file
[http://www.biomedcentral.com/content/supplementary/1757-5036-1-6-S2.pdf]

### Additional File 3
*FRET controls.* Fluorescence emission of AChR-only and PyPC-only samples as a function of temperature. Fig. 3 and experimental details.
Click here for file
[http://www.biomedcentral.com/content/supplementary/1757-5036-1-6-S3.pdf]

## Acknowledgements
This work was supported in part by grants PICT 01-12790 and 5-20155 from FONCYT, PIP No. 6367 from the Argentinian Scientific Research Council (CONICET) and PGI No. 24/B135 from Universidad Nacional del Sur, Argentina, to FJB.